\documentclass[twocolumn,journal]{IEEEtran}

\usepackage{graphicx}
\usepackage{amsmath}
\usepackage{amssymb}
\usepackage{comment}
\usepackage{tabu}
\usepackage{amsthm}
\usepackage{cite}
\usepackage{float}
\usepackage{color}
\usepackage{balance}

\begin{document}
\title{Sparsity-Cognizant Multiple-Access Schemes for Large Wireless Networks With Node Buffers}
\author{Ahmed El Shafie$^\dagger$, Naofal Al-Dhahir$^\dagger$ and Ridha Hamila$^*$\\
\begin{tabular}{c}
 $^\dagger$Electrical Engineering Dept., University of Texas at Dallas, USA. \\
  $^*$Electrical Engineering Dept., Qatar University, Doha, Qatar.
\end{tabular}
\thanks{This paper was made possible by NPRP grant number 6-070-2-024 from the Qatar National Research Fund (a member of Qatar Foundation). The statements made herein are solely the responsibility of the authors.}\thanks{This paper was published in the IEEE 12th International Conference on Mobile Ad Hoc and Sensor Systems 2015.}
}

\date{}
\maketitle
\thispagestyle{empty}
\pagestyle{empty}
\begin{abstract}
This paper proposes efficient multiple-access schemes for large wireless networks based on the transmitters' buffer state information and their transceivers' duplex transmission capability. First, we investigate the case of half-duplex nodes where a node can either transmit or receive in a given time instant. In this case, for a given frame, the transmitters send their buffer states to the destination which assigns the available time duration in the frame for data transmission among the transmitters based on their buffer state information. The network is said to be naturally sparse if the number of nonempty-queue transmitters in a given frame is much smaller than the number of users, which is the case when the arrival rates to the queues are very small and the number of users is large. If the network is not naturally sparse, we design the user requests to be sparse such that only few requests are sent to the destination. We refer to the detected nonempty-queue transmitters in a given frame as \emph{frame owners}. Our design goal is to minimize the nodes' total transmit power in a given frame. In the case of unslotted-time data transmission, the optimization problem is shown to be a convex optimization program. We propose an approximate formulation to simplify the problem and obtain a closed-form expression for the assigned time durations to the nodes. The solution of the approximate optimization problem demonstrates that the time duration assigned to a node in the set of \emph{frame owners} is the ratio of the square-root of the buffer occupancy of that node to the sum of the square-roots of each occupancy of all the \emph{frame owners}. We then investigate the slotted-time data transmission scenario, where the time durations assigned for data transmission are slotted. In addition, we show that the full-duplex capability of a node increases the data transmission portion of the frame and enables a distributed implementation of the proposed schemes. Our numerical results demonstrate that the proposed schemes achieve higher average bits per unit power than the fixed-assignment scheme where each node is assigned a predetermined fraction of the frame duration.
\end{abstract}
\section{Introduction}
Motivated by the goal of increasing spectrum efficiency, we propose a set of efficient multiple-access schemes for large wireless networks. When the number of transmitters is large, the communication between the nodes consumes most of the available time and spectrum resources. Therefore, designing an efficient scheme for such scenarios is of great interest.

The compressive sensing (CS) paradigm is an efficient approach that enables the receiver to reconstruct the original sparse signal from few noisy
observations \cite{1580791,1614066}. The authors in \cite{bajwa2006compressive,bajwa2007joint} use phase-coherent analog transmission of randomly-weighted data from sensor nodes to the fusion center. The additive property of the multiple-access channel enables projections
of data onto an appropriate basis at the fusion center.
 The authors of \cite{fazel2011random} investigated a random-access CS-aided scheme
for underwater sensor networks. Their goal was to
design a power-efficient random data collection scheme.
A CS-aided medium access control (CS-MAC)
scheme is proposed in  \cite{lin2012compressive} where the access point (AP) allocates a random
sequence to each user.
All user requests for gaining uplink
transmissions access are sent simultaneously in a
synchronous manner. The authors of \cite{6554680} studied the asynchronous scenario of \cite{lin2012compressive}. The data arrival dynamics and queue states were not considered in \cite{lin2012compressive,6554680}.


In this paper, we design efficient sparsity-aware multiple-access schemes for large wireless networks with node buffers. The contributions of this paper are summarized as follows. We use the CS techniques to realize our proposed MAC schemes. We investigate the scenarios when the system is naturally sparse and when it is non-sparse. We investigate the case of unslotted-time data transmission. Based on the buffer state information (BSI) at the transmitters, the destination assigns equivalent time durations to each transmitter. We investigate the case of slotted-time data transmission system, where data are sent to the destination in a slotted manner, and design the system and its sparsity level based on the available number of data time slots per frame. To avoid wasting the data time slots, we propose two solutions for the case when the number of detected users per frame is lower than the available data time slots. We investigate the cases of half-/full-duplex nodes and show that the full-duplex capability enables the nodes to implement the proposed schemes in a distributed manner and reduces the needed time for control signals overhead.

\emph{\underline{Notation:}}  $\mathbb{C}^\mathcal{M}$ denotes the set of complex vectors of size $\mathcal{M}$. $|\cdot|$ denotes the absolute value. The
operators $\|\cdot\|_0$, $\|\cdot\|_1$, $\|\cdot\|_2$ denote $\ell_0$-norm, $\ell_1$-norm, and
$\ell_2$-norm of a vector, respectively. In addition, $(\cdot)^\dagger$, $(\cdot)^{-1}$, and
$(\cdot)^\mathbb{T}$ denote the Hermitian (i.e. complex-conjugate transpose),
inverse, and transposition operations, respectively. A summary of the key variables adopted in this paper is provided in Table \ref{table1}.

The remainder of this paper is organized as follows. In the next section, we give a brief introduction to CS. In Section~\ref{sys}, we discuss the system model adopted in this paper and describe our proposed schemes. The simulation results are provided in Section \ref{sim}. We conclude the paper in Section~\ref{conc}.
 \begin{table}
\renewcommand{\arraystretch}{1}
\begin{center}
  {\tabulinesep=0.7mm \begin{tabu}{ c |l  }
    \hline\hline
    Symbol & Description \\[5pt]\hline
      $\mathcal{N}$ & {\footnotesize Number of users} \\[5pt]\hline
       $W$ and $T$& {\footnotesize Channel bandwidth and frame duration} \\[5pt]\hline
      {\footnotesize $\tilde K$} & {\footnotesize Number of frame owners} \\[5pt]\hline
        $\mathcal{M}$ & {\footnotesize Number of measurements needed to know the buffer states} \\[5pt]\hline
        $\mathcal{A}$ & {\footnotesize Measurement matrix} \\[5pt]\hline
         $\mathcal{S}$ & {\footnotesize System sparsity level} \\[5pt]\hline
           $\mathcal{L}$ & {\footnotesize Maximum capacity of buffer $Q_k$ in packets} \\[5pt]\hline
            $b$ & {\footnotesize Number of bits in a data packet} \\[5pt]\hline
             $\lambda_k$ & {\footnotesize Mean packet arrival rate at $Q_k$} \\[5pt]\hline
         $P_k$  & {\footnotesize Average transmit power of node $k$ in Watts/Hz} \\ & {\footnotesize during the current frame} \\[5pt]\hline
             $\kappa$ & {\footnotesize Noise power spectral density in Watts/Hz} \\[5pt]\hline
    $B^\ell_k$ & {\footnotesize Number of bits announced to be stored at queue $Q_k$}\\  & {\footnotesize when the user sends a request during frame $\ell$} \\[5pt]\hline
     $B^\ell_k=0$ & {\footnotesize Number of bits that user $k$ wants to send is zero}\\  & {\footnotesize when it has no data or decides not to send a request} \\[5pt]\hline
      $\mathcal{C}_k$ & {\footnotesize Capacity of link $k$} \\[5pt]\hline
       $T_k$ & {\footnotesize Time needed to transmit the $B_k$ bits at head of $Q_k$}\\  & {\footnotesize in case of unslotted-time data transmission} \\[5pt]\hline
        $\alpha$ & {\footnotesize Request probability} \\[5pt]\hline
         $D$ & {\footnotesize Number of data time slots per frame}\\  & {\footnotesize in case of slotted-time data transmission} \\[5pt]\hline
          $n_k$ & {\footnotesize Number of data time slots assigned to user $k$}\\   & {\footnotesize in case of slotted-time data transmission}
          \\[5pt]\hline
            $T_{\rm s}$ & {\footnotesize Data time slot duration}\\  & {in case of slotted-time data transmission} \\[5pt]\hline
\end{tabu}}
\end{center}
\caption{List of Key Variables.}
\label{table1}
\vspace{-0.9cm}
\end{table}
\section{CS Background}
\label{cs}
In this section, we introduce the CS principles. A comprehensive treatment of the subject is given in \cite{sensing} and the references therein.
CS approximations were originally stimulated by the idea that natural signals are often
sparse in some domain.
That is, a signal can be represented by
a few significant coefficients in some representative domain. For instance, natural images are sparse in the
frequency domain.
On one hand, traditional compression can exploit
this sparsity to reduce the signal size, however, this requires statistical analysis
of the entire signal to find the significant coefficients. On the other hand, the CS approach removes this requirement.
For a given sparse signal of
sparsity level $\mathcal{S}$, i.e., it has at most $\mathcal{S}$ significant (nonzero) coefficients, we can
capture the $\mathcal{S}$ coefficients without the need to collect the complete data set of some large size $\mathcal{N}\gg S$.



Consider the system of equations $\mathbf{Y}=\mathcal{A} \mathbf{X}+\mathbf{Z}$ where $\mathbf{Y}\in \mathbb{C}^\mathcal{M}$
is a known measurement vector, $\mathbf{X}\in \mathbb{C}^\mathcal{N}$ is an unknown vector,
$\mathcal{A}$ denotes the $\mathcal{M} \times \mathcal{N}$ measurement matrix, and $\mathbf{Z} \in \mathbb{C}^\mathcal{M}$ is a
bounded noise (error) vector. If the measurement matrix $\mathcal{A}$ satisfies the so-called restricted isometry property (RIP) condition \cite{sensing,candes2006stable}, the sparse vector $\mathbf{X}$ can be recovered with a number of measurements on the order of $\mathcal{S} \log_{10} (\mathcal{N}/\mathcal{S}) \ll \mathcal{N}$ measurements, where $\mathcal{S}$ is the sparsity level of $\mathbf{X}$.
 To obtain the sparsest solution
to this system of equations, the following problem is solved
\begin{eqnarray}
\small \begin{split} \small
    \underset{{\mathbf{X}}}{\min} & \ \ \ \ \|{\mathbf{X}}\|_0,\ {\rm s.t.}   \ \ \ \ \|{\mathbf{Y}}-\mathcal{A}{\mathbf{X}}\|_2^2\le \delta
     \end{split}
\end{eqnarray}
where $\delta$ is chosen such that it bounds the amount of noise in the measurements and $\|{\mathbf{X}}\|_0$ is the
number of nonzero entries in $\mathbf{X}$. In general, finding
the optimal solution to this problem is not computationally
efficient. Hence, two main approaches have been proposed
in the literature to efficiently compute a sparse suboptimal solution to
this system of equations; specifically, $\ell_1$-norm minimization
and greedy algorithms. As shown in the CS literature, we can recover the sparse signal
$\mathbf{X}$ using the following constrained $\ell_1$-norm minimization problem \cite{candes2006stable}
\begin{eqnarray}
\small \begin{split} \small
    \underset{{\mathbf{X}}}{\min} & \ \ \ \ \|{\mathbf{X}}\|_1\ {\rm s.t.}   \ \ \ \ \|{\mathbf{Y}}-\mathcal{A}{\mathbf{X}}\|_2^2\le \delta
    \label{ff}
     \end{split}
\end{eqnarray}
However,
the solution of the optimization problem is not exactly sparse because many nonzero entries with nonsignificant values can typically appear in ˜$\mathbf{X}$. An additional heuristic optimization steps can be applied to
enforce a finite number of nonzero entries as in \cite{baran2010linear}. Alternatively,
the greedy algorithms provide more control on the set of
nonzero elements whose indices and values are determined iteratively.
We describe the orthogonal matching pursuit (OMP)
algorithm \cite{tropp2007signal} as one of the widely-used greedy algorithms. OMP takes $\mathbf{Y}$, $\mathcal{A}$, and a certain stopping criterion as its inputs
and computes a sparse solution $\hat{\mathbf{X}}$ for the unknown vector
$\mathbf{X}$ as its output. Hence, we denote the OMP operation by
$\hat{\mathbf{X}} = {\rm OMP} (\mathbf{Y}, \mathcal{A},{\rm stopping \ criterion})$. The stopping criterion
can be a predefined sparsity level (number of nonzero entries) of $\mathbf{X}$ or an upper bound on the norm of the residual error
term  $\|\mathbf{Y}-\mathcal{A}\mathbf{X}\|^2_2$.



\section{System Model and Proposed Schemes}
\label{sys}
\begin{figure}
  \centering
  \includegraphics[width=1\columnwidth]{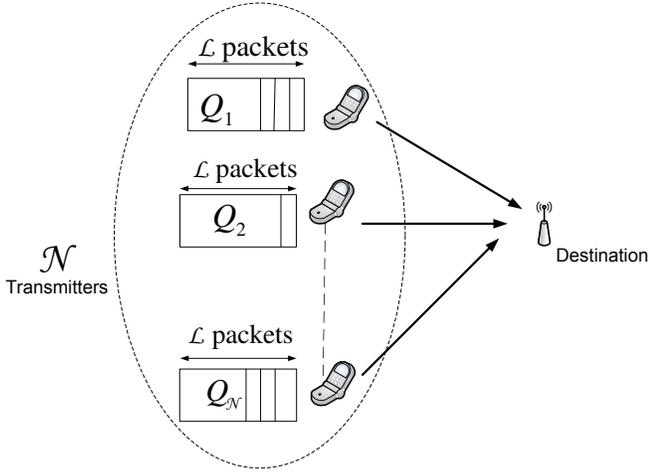}\\
  \caption{The system model adopted in this paper.}
  \label{fig0}
  \end{figure}

We assume an uplink scenario where a set of buffered transmitters wishes to communicate with a common destination (base-station) as shown in Fig. \ref{fig0}. The set of transmitting nodes is labeled as $1,2,\dots,\mathcal{N}$ and the destination is denoted by ${\rm d}$. We consider buffered nodes where node $k$ has a finite-length buffer (queue), denoted by $Q_k$, whose maximum size is $\mathcal{L}$ packets. The size of a packet is $b$ bits. Each source node transmits its buffer state (i.e. the number of packets stored in its queue) to the destination. We assume Gaussian channels\footnote{For the case of fading channels, the only difference is that the transmit powers will be scaled by the fading coefficients. That is, if the power used by user $k$ is $P_k$ without fading, the power with fading is $\tilde P_k=P_k/|h_k|^2$, where $h_k$ is the fading coefficient between node $k$ and the destination. To avoid very large transmit powers specially at very low fading gain, i.e., very low $|h_k|^2$, we suggest two modifications to the system: (1) we make the users with very low fading gains OFF/silent during the current frame. This increases the system sparsity level. We can adjust the request probability based on the fading value, (2) we put a threshold on the maximum transmit power such that the users either transmit data corresponding to the maximum transmit power or remain silent if the needed power is higher than the maximum level.} where a received signal is corrupted by an additive white Gaussian noise (AWGN) with zero mean and variance $\kappa$ Watts/Hz. The time is partitioned into frames each with duration $T_f$ time units.

Assuming that the average transmit power of transmitter $k$ per unit frequency is $P_k$ Watts/Hz, the channel capacity of link $k$ (i.e. the link connecting transmitter $k$ and the base-station) is then given by
 \begin{equation}
\mathcal{C}_k=\log_2\left(1+\frac{P_k}{\kappa}\right)
\end{equation}
If transmitter $k$ wishes to send $B_k$ bits, the time required for such transmission, given a channel bandwidth of $W$ Hz, is given by
\begin{equation}
\label{gogo}
T_k=\frac{B_k}{W \mathcal{C}_k}= \frac{B_k}{ W \log_2\left(1+\frac{P_k}{\kappa}\right)}
\end{equation}
Each transmitter sends its buffer state as a single signal with an amplitude corresponding to its value in packets. The used constellation points are $0,1,2,\dots,\mathcal{L}$, where $\mathcal{L}$ is the maximum number of data packets that can be stored at any of the queues. The time needed for bits transmission, $T_k$, is obtained at the base-station upon demodulating the signal representing the number of bits stored in $Q_k$ using the expression in (\ref{gogo}). If the system is sparse or can be made sparse using a request probability as will be explained later, the base-station then knows the time duration needed for each node using CS approximations. In this case, the communication of the buffer state information requires only $\mathcal{M}\ll \mathcal{N}$ measurements. If the system can not be made sparse, $\mathcal{N}$ orthogonal signals should be used by the transmitters for their communication with the base-station.

If the system is sparse, the buffer state signal sent by a user is modulated (multiplied) by a sequence (e.g. Bernoulli, Gaussian, or Gold codes) of length $\mathcal{M}$ bits. The sequences of all transmitters represent the measurement matrix which is used by the destination to resolve the transmitter requests. The transmitters request vector received at the destination is
\begin{equation}
\mathbf{Y}^{\rm d}= \mathcal{A} \mathbf{X}+\mathbf{Z}^{\rm d}
\end{equation}
where $\mathbf{Y}^{\rm d}$ is the measurement vector received at the destination, $\mathbf{X}=[x_1,x_2,\dots,x_\mathcal{N}]^\mathbb{T}$ is the user buffer states vector with $x_k$ representing the buffer state of transmitter $k$, $\mathcal{A}$ is the measurement matrix adopted by the system, and $\mathbf{Z}^{\rm d}$ is the additive noise vector at the destination whose size is $\mathcal{M}\time 1$. We assume that the transmitters send their requests with sufficient powers such that the requests are decoded perfectly at the destination as long as the requests are less than or equal to the system sparsity level. Thus, if the number of requests is less than or equal to the sparsity level, the nonzero components are decodable with probability $1$.

It should be mentioned here that the output of the CS solver is used as follows. The location of the nonzero element represents the transmitter index while the demodulated value represents the number of packets (or bits) stored at the sender's buffer. Since the output may contain errors, we assume that the output for the $k$th user is $\hat{B}_k=b \hat{x}_k$ bits (i.e. estimated value of $B_k=b x_k$ bits), where $\hat{x}_k$ is the output corresponding to the $k$th user in packets and $b$ is the packets size in bits. Note that $\hat{x}_k$ must be one of the constellation points used to represent $x_k$. If user $k$ has no data to send or decides not to access the channel (i.e. decides not to send a request which occurs with probability $1-\alpha$), $B_k=0$ and, hence, $T_k=P_k=0$. The corresponding time for sending the $\hat{B}_k$ bits is $\hat{T}_k=\frac{\hat{B}_k}{W \mathcal{C}_k}$. The time needed by all transmitters, from the base-station viewpoint, is $T=\sum_{k=1}^\mathcal{N} \hat{T}_k$. In a given frame, we refer to the detected nonempty-queue transmitters as the \emph{frame owners}. The destination sends the time durations to the \emph{frame owners}.
 If the received average signal-to-noise ratio (SNR) at the destination is high, we can assume that the destination knows the buffer states at the transmitters perfectly. Hence, $\hat{B}_k=B_k$ and $\hat{T}_k=T_k$.
%

Let $\mathbb{A}^\ell_k$ denote the number of packets that arrive at the buffer of transmitter $k$ at the end of frame $\ell\in\{1,2,\dots\}$, which can follow any distribution, with average rate $\lambda_k$ packets per frame (packets/frame). At the end of frame $\ell$ (i.e. beginning of frame $\ell+1$), the number of packets remaining at the $k$th transmitter's queue is
\begin{equation}
Q^{\ell+1}_k\leftarrow \min\left\{Q^{\ell}_k- \min\{x_k,\hat{x}_k\}+ \mathbb{A}^\ell_k,\mathcal{L}\right\}
\end{equation}
where $Q^{\ell}_k$ is the buffer size at the beginning of frame $\ell$ and $\mathcal{L}$ is the buffer maximum capacity in packets. Since the queue states are detected accurately at the destination, $\min\{x_k,\hat{x}_k\}=\hat{x}_k=x_k$.
It should be noted that the random characteristics of the arrivals at transmitters increase the sparsity of the system. This means that the sporadic nature of the arrivals increases the possibility of having many empty-queue transmitters and hence lower requests. Assume Bernoulli arrivals with mean $\lambda_k=\lambda$ for all queues,\footnote{If all queues have different average arrival rates, we replace $\lambda$ with the maximum among all arrival rates. That is, $\lambda=\max\{\lambda_1,\lambda_2,\dots,\lambda_\mathcal{N}\}$.} where $\lambda$ is the probability of having an arrived packet in a given frame. In addition, assume the case when all queues are empty in a given frame. The probability of having \emph{at least} $K$ arrived packets in the system during the next frame is
\begin{equation}
\label{sparsity2}
\sum_{j=K}^\mathcal{N} \left(\!\begin{array}{c}
\mathcal{N} \\
  j
\end{array}\right) \lambda^ j (1-\lambda)^{\mathcal{N}-j}
\end{equation}
where $\Big(\!\begin{array}{c}
\mathcal{N} \\
  j
\end{array}\!\Big)$ denotes $\mathcal{N}$ choose $j$. The system is sparse with sparsity level $\mathcal{S}=K-1$ if this probability is very small which is the case as long as $\lambda$ is very small and $\mathcal{N}$ is large. In case of other arrival distributions (i.e. non-Bernoulli arrivals), $\lambda$ in Eqn. (\ref{sparsity2}) is replaced with the probability that the queue has \emph{at least} one packet and $1-\lambda$ is the probability of having no packets (i.e. the queue is empty). If the system is sparse due to large $\mathcal{N}$ and very small $\lambda$, all users can attempt to send their queue states to the destination in every frame.

When the system is not naturally sparse, exploiting the request probability is necessary to ensure sparsity of the system. In this case, the system sparsity condition is
\begin{equation}
\label{sparsity}
\sum_{j=\mathcal{S}+1}^\mathcal{N} \left(\begin{array}{c}
\mathcal{N} \\
  j
\end{array}\right) \alpha^ j \left(1-\alpha\right)^{\mathcal{N}-j} \le \epsilon
\end{equation}
where $\mathcal{S}$ is the sparsity level and $\epsilon$ is a very small positive number. This condition is based on the worst case scenario when all transmitters have data to send. Designing the system based on this criterion guarantees the sparsity of the system in all other cases of queue occupancies (i.e. when a set of nodes is empty and the other sets are not empty). The optimal $\alpha$ is the highest feasible value of $\alpha$ that satisfies the sparsity-level constraint in (\ref{sparsity}). This is because as $\alpha$ increases, the number of served users increases. However, this also decreases the ability of the destination to decode the users request vector using only $\mathcal{M}\ll \mathcal{N}$ measurements.

The sequences used by the transmitters can be seen as random-access codes that are used to distinguish between the users at the destination.
Our goal is to accommodate as many transmitters as possible; hence, we increase the length of the sequences. However,
longer sequences lead to longer communication time, i.e., longer time needed to capture more observations at the destination to determine all the nonempty-queue transmitters,
and consequently a lower data transmission time is achieved.


\subsection{Unslotted-Time Data Transmission}
\begin{figure}
  \centering
  \includegraphics[width=0.6\columnwidth]{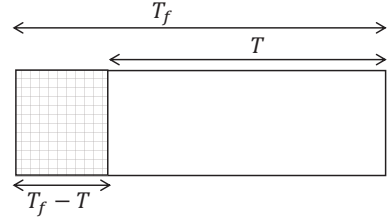}\\
  \caption{Frame structure for the unslotted-time data transmission.}
  \label{unslot}
  \end{figure}

In this scenario, we assume that the communication between the transmitters and the destination is divided into three stages. In the first stage, the transmitters send their buffer states to the destination. In the second stage, the destination assigns the transmission time to each user. In the third stage, the \emph{frame owners} transmit their data to destination. The frame structure under this scenario is shown in Fig. \ref{unslot}.
Our proposed communication protocol during a frame for the unslotted-time data transmission scenario is described as follows:
\begin{itemize}
\item If the system is naturally sparse, i.e., satisfies the condition in Eqn. (\ref{sparsity2}), the users send their BSI to the destination in each frame using CS approximations without using the request probability.
\item If the system is not naturally sparse, the transmitters probabilistically send their requests to the destination. The access probability is $0\le \alpha\le1$.
\item Upon decoding the buffer state vector, the destination sends feedback signals to the transmitters which contain the time assignments among the transmitters. This can be implemented as follows. The destination sends an acknowledgement (ACK) signal with the identification (ID) of the transmitter which is assigned to the next dedicated time duration. Then, the length of the time duration assigned to that user is broadcasted. Each of these activities requires $1/W$ seconds to be implemented; hence, the total time spent in identifying the next transmitter and its transmission time declaration is $2/W$ time units. We assume that the errors in decoding these values at all nodes are negligible.\footnote{In practice, the probability of decoding error is very small. For example, using $16$-QAM, the bit error rate is $10^{-4}$ when the SNR is $12$ dB.}
\item Each transmitter sends its data during its assigned transmission time.
\end{itemize}
After sending the transmitter requests and the feedback signals, the remaining time for data transmission is $T=T_f-\frac{\mathcal{M}}{W}-2\frac{\tilde K}{W}=T_f-\frac{\mathcal{M}+2 \tilde K}{W}$, where $T_f$ is the frame duration, $\tilde K\le \mathcal{S}$ is the total number of nonempty-queue transmitters during the current frame (i.e. number of \emph{frame owners}), and $2 \tilde K/W$ is the total time used to inform the \emph{frame owners} about their dedicated time durations.

The optimization problem that minimizes the total transmit power of nodes under the constraint that the sum of the users' transmission times is at most $T$ is stated as follows:
\begin{eqnarray}
\small \begin{split} \small
\label{opti1}
   \underset{0\le P_k\le \infty, \forall k }{\min}  & \,\,\,\,\  \sum_{k=1}^{\mathcal{N}} P_k\\ {\rm s.t.}   & \,\,\,\,\
\sum_{k=1}^{\mathcal{N}} T_k= \sum_{k=1}^{\mathcal{N}}\frac{\hat{B}_k}{ W \log_2\left(1+\frac{P_k}{\kappa}\right)}\le T
     \end{split}
\end{eqnarray}
The objective function is linear in $P_k$. The second derivative of the term $T_k=\frac{\hat{B}_k}{ W \log_2\left(1+\frac{P_k}{\kappa}\right)}$ is
\begin{eqnarray}
\small \begin{split} \small
\frac{\delta^2 T_k}{\delta P_k^2}=\frac{\hat{B}_k \ln(2) \left(\ln(1+\frac{P_k}{\kappa})+2\right)}{W (\kappa+P_k)^2 \ln^3(1+\frac{P_k}{\kappa})} \ge 0
     \end{split}
\end{eqnarray}
The second derivative is always positive for $P_k\ge 0$; hence, $T_k$ is convex in $P_k$. Since the positive weighted sum of convex functions is also convex, the constraint $\sum_{k=1}^{\mathcal{N}} T_k$ is convex.
Since the objective function is linear and the constraint is convex, our optimization problem is a convex program and can be solved efficiently using any convex optimization solver \cite{boyed}. To gain further insights and derive closed-form expressions for the time assignments, we seek an approximation for the optimization problem in (\ref{opti1}).
%
Letting $\frac{1}{T_k}=\frac{\log_2(1+\frac{P_k}{\kappa})}{a_k}$, where $a_k=\frac{\hat{B}_k}{ W }$, $P_k=\kappa(2^{\frac{a_k}{T_k}}-1)$. Thus, the optimization problem is rewritten as
\begin{eqnarray}
\small \begin{split} \small
   \underset{0\le T_k\le \infty, \forall k }{\min}  & \,\,\,\,\  \sum_{k=1}^{\mathcal{N}}\kappa(2^{\frac{a_k}{T_k}}-1)\\ {\rm s.t.}   & \,\,\,\,\
 \sum_{k=1}^{\mathcal{N}}T_k\le T
     \end{split}
\end{eqnarray}
Eliminating the constant terms from the objective function, the optimization problem is rewritten as
\begin{eqnarray}
\small \begin{split} \small
\label{sa3er}
   \underset{0\le T_k\le \infty, \forall k }{\min}  & \,\,\,\,\  \sum_{k=1}^{\mathcal{N}}2^{\frac{a_k}{T_k}} \\ {\rm s.t.}   & \,\,\,\,\
 \sum_{k=1}^{\mathcal{N}}T_k\le T
     \end{split}
\end{eqnarray}


 Taking the logarithm of base $2$ for the objective function, and using Jensen's inequality \cite{kuczma2009introduction}, the objective function of (\ref{sa3er}) can be upper bounded as follows:
 \begin{eqnarray}
\small \begin{split} \small
\log_2 \left (\sum_{k=1}^{\mathcal{N}}2^{\frac{a_k}{T_k}}\right) \le  \sum_{k=1}^{\mathcal{N}}\log_2 \left(2^{\frac{a_k}{T_k}}\right)=\sum_{k=1}^{\mathcal{N}} \left({\frac{a_k}{T_k}}\right)
     \end{split}
\end{eqnarray}
 The approximate optimization problem is stated as follows:
 \begin{eqnarray}
\small \begin{split} \small
\label{romanna}
   \underset{0\le T_k\le \infty, \forall k }{\min}  & \,\,\,\,\  \sum_{k=1}^{\mathcal{N}}{\frac{a_k}{T_k}} \\ {\rm s.t.}   & \,\,\,\,\
 \sum_{k=1}^{\mathcal{N}}T_k\le T
     \end{split}
\end{eqnarray}
This approximation is tight when $a_k$ is very small; which is the case when the buffer capacity is much less than the bandwidth, i.e., $\mathcal{L}\ll W$. Noting that increasing the time assigned for data transmission decreases the objective function, the constraint in (\ref{romanna}) holds with equality. That is, $\sum_{k=1}^{\mathcal{N}}T_k= T$. Using the Lagrange multipliers, we get
 \begin{eqnarray}
\small \begin{split} \small
\mathbf{L}= \sum_{k=1}^{\mathcal{N}}{\frac{a_k}{T_k}}+\nu \sum_{k=1}^{\mathcal{N}}T_k
     \end{split}
\end{eqnarray}
  where $\mathbf{L}$ is the Lagrange function and $\nu$ is the Lagrange multiplier. Taking the first derivative of $\mathbf{L}$ with respect to $T_k$, we get
 \begin{eqnarray}
\small \begin{split} \small
\frac{\delta \mathbf{L}}{\delta T_k}= -{\frac{a_k}{T^2_k}}+\nu
     \end{split}
\end{eqnarray}
Equating this value to zero and setting $T_k=T_k^*$, we get
\begin{eqnarray}
\small \begin{split} \small
\label{gogp}
 T_k^*= \sqrt{\frac{a_k}{\nu}}
     \end{split}
\end{eqnarray}
Using the constraint that $\sum_{k=1}^\mathcal{N} T_k=T$, we calculate $\nu$ as follows:
\begin{eqnarray}
\small \begin{split} \small
 \sum_{k=1}^\mathcal{N} T_k^*=T=  \sum_{k=1}^\mathcal{N}  \sqrt{\frac{a_k}{\nu}}
     \end{split}
\end{eqnarray}
Thus,
\begin{eqnarray}
\small \begin{split} \small
\label{nununu}
 \sqrt{\nu}=\sum_{k=1}^\mathcal{N} T_k^*=  \sum_{k=1}^\mathcal{N}  \frac{\sqrt{{a_k}}}{T}
     \end{split}
\end{eqnarray}
Substituting (\ref{nununu}) into (\ref{gogp}), the optimal $T_k$ that minimizes the approximate optimization problem is given by
\begin{eqnarray}
\small \begin{split} \small
\label{gogp2}
 T_k^*= T \frac{\sqrt{{a_k}}}{\sum_{k=1}^\mathcal{N}  {\sqrt{{a_k}}}}=T \frac{\sqrt{{\hat{B}_k}}}{\sum_{k=1}^\mathcal{N}  {\sqrt{{\hat{B}_k}}}}
     \end{split}
\end{eqnarray}
The solution of the approximate formulation illustrates the fact that the users with the higher queue occupancies should be assigned longer time durations. It should be noted here that the achievable rate under this scheme is $\sum_{k=1}^\mathcal{N} B_k$ bits/frame. If all \emph{frame owners} have the same queue occupancy, the optimal solution will be $T/\tilde K$, i.e., splitting the remaining time in the frame, $T$, equally among the \emph{frame owners}.

Using the optimal solution in (\ref{gogp2}), the total transmit power in this case is
\begin{eqnarray}
\small \begin{split} \small
 P_{\rm upper}=\sum_{k=1}^{\mathcal{N}} P_k =\sum_{k=1}^\mathcal{N} \kappa(2^{\sqrt{B_k}\sum_{k=1}^\mathcal{N} \frac{\sqrt{B_k}}{TW}}-1)
     \end{split}
\end{eqnarray}
The value $P_{\rm upper}$ is an upper bound on the total transmit power. Since the queues are limited in size, assuming that all queues are full of data packets, the maximum value of $P_{\rm upper}$, denoted by $P^{\max}_{\rm upper}$, is given by
 \begin{eqnarray}
\small \begin{split} \small
 P^{\max}_{\rm upper}=\sum_{k=1}^{\mathcal{N}} P_k = \kappa \tilde K (2^{\tilde K\frac{\mathcal{L} b}{TW}}-1)
     \end{split}
\end{eqnarray}
 where $\tilde K$ is the number of active transmitters.

If we assume a {\it virtual} buffer with a total number of bits $\overline{B}=\sum_{k=1}^\mathcal{N} B_k$ in frame $\ell\in\{1,2,\dots\}$, the power needed to send the $\overline{B}$ bits during the data transmission time of the frame, $T$, is given by
\begin{eqnarray}
\small \begin{split} \small
  P_{\rm lower}= (2^{\frac{\overline{B}}{WT}}-1)\kappa
     \end{split}
\end{eqnarray}
where $P_{\rm lower}$ is the minimum power level needed to send all bits. It should be noted that this scenario, where a virtual buffer maintains a total number of bits equal to the number of bits in the \emph{frame owners}' buffers, is an upper bound on what can be achieved in the original scenario where we have a set of nodes (i.e. \emph{frame owners}) each of which has its own data packets stored at its own buffer and wishes to send them to its destination. Hence, the transmit power in this virtual system is a lower bound on the transmit power of the original scenario. To demonstrate this point, we study the following example. Assume that we have only two \emph{frame owners}, say nodes $1$ and $2$, where each of them has only one packet of size $b$ bits, i.e., $B_1=B_2=b$ bits. In this case, and since we have only two nodes with equal buffer sizes, the time duration $T$ is divided equally between them. Hence, the total transmit power is
 \begin{eqnarray}
\small \begin{split} \small
  P_1+P_2= 2 (2^{\frac{b}{WT/2}}-1)\kappa=2 (2^{\frac{2b}{WT}}-1)\kappa
     \end{split}
\end{eqnarray}
If we assume that the two packets are stored in a single buffer (i.e. virtual buffer of a total size equal to $2b$ bits), the needed power to transmit $\overline{B}=2b$ bits over $T$ time units is
 \begin{eqnarray}
\small \begin{split} \small
 P_{\rm lower}= (2^{\frac{2b}{WT}}-1)\kappa=\frac{1}{2} (P_1+P_2)
     \end{split}
\end{eqnarray}

Based on the above argument, the total transmit power is lower bounded as
\begin{eqnarray}
\small \begin{split} \small
   \sum_{k=1}^\mathcal{N} P_k \ge P_{\rm lower}= (2^{\frac{\overline{B}}{WT}}-1)\kappa
     \end{split}
\end{eqnarray}

The total transmit power obtained from solving the original optimization problem in (\ref{opti1}) may take any value between $P_{\rm lower}$ and $P_{\rm upper}$. That is, $P_{\rm lower}\le\sum_{k=1}^\mathcal{N} P_k\le P_{\rm upper}$.

\subsection{Slotted-Time Data Transmission}
In this scenario, we assume that the communication between the transmitters and the destination is divided into three stages. In the first stage, the transmitters attempt to send their buffer states to the destination. In the second stage, upon getting the compressed signal of the buffer states, the transmitters are assigned to a set of $D$ data time slots each with duration $T_{\rm s}$ time units. In the third stage, the users assigned to the available data time slots send their data to the destination. The frame structure under this scenario is shown in Fig. \ref{slotted}. The assignment of time slots to transmitters depends on the number of \emph{frame owners} and the number of data time slots per frame. That is,
 \begin{itemize}
 \item If the system is designed such that the number of requests is always lower than or equal to the number of data time slots, i.e., $\mathcal{S}\le D$, then all decoded requests will be satisfied, i.e., each transmitter gets one time slot. However, in this case, we may have some unused data time slots if the number of \emph{frame owners} is lower than the number of data time slots, $D$.
     \item If the number of requests (i.e. \emph{frame owners}) exceeds the number of available data time slots, the $D$ \emph{frame owners} with highest queues occupancy will be selected for data transmission.
         \end{itemize}
          In this system, the destination communicates with the users during the control signals exchange in an unslotted manner. Note that increasing $\mathcal{S}$ allows selection of more transmitters such that the users with high queues occupancy can be detected adequately. The access probability can be increased and hence the detected users (requests). Adding one additional user to the problem results in time loss that increases the sparsity level to $\mathcal{S}+1$ instead of $\mathcal{S}$. Since the users probabilistically send their BSI to the destination in each frame, we may have a number of \emph{frame owners} less than $D$ as explained in the above-mentioned cases. Hence, some time slots may be wasted. We propose two solutions to this problem.
 \begin{enumerate}
 \item Since losing a time slot with duration $T_{\rm s}$ can be worse than losing a time duration of $m/W$ when $T_{\rm s}>m/W$, where $m$ is a positive integer and $m/W$ is the time needed to increase the sparsity level from $\mathcal{S}$ to $\mathcal{S}+n$ for accepting $n$ additional users per frame, we can increase the number of users per frame by increasing the request probability, $\alpha$. However, since we have only $D$ data time slots per frame, more requests result in less time for data transmission per time slot, i.e., the data time slot duration, $T_{\rm s}$, decreases as number of requests increases. Since the number of \emph{frame owners} can be less than $D$, some time slots may be unutilized in a given frame. Hence, the cardinality of the set of users that will use the available time slots in a given frame is $\mathcal{T}\le D$, where $\mathcal{T}\le \tilde K$. In this approach, the destination does not need to send the transmission times and instead it sends the value of $\mathcal{T}$ in the current frame and broadcasts the IDs of the users that will use the available time slots. Accordingly, the communication of the control signals between the destination and the users requires $1/W+\mathcal{T} /W$ time units, where $1/W$ is the time consumed for announcing the value of $\mathcal{T}$ and $\mathcal{T} /W$ is the time needed for announcing the IDs of the group of users that will use the data time slots. Consequently, the data time slot duration is $ T_{\rm s}=\max\{\frac{T-{\mathcal{M}/W - \mathcal{T}/W-1/W}}{D},0\}$. Note that the control signal exchange takes place over the time interval from instant $0$ to instant $\mathcal{M}/W + \mathcal{T} /W$ and data transmission takes place over the time interval from instant ${\mathcal{M}/W + \mathcal{T} /W+1/W}$ to instant $ {\mathcal{M}/W + \mathcal{T}/W+1/W} +D T_{\rm s}$. Using this approach, the complexity of the system is reduced and the time slots loss and time durations broadcasting is reduced as well. One may optimize both the number of time slots per frame, $D$, and the system sparsity level, $\mathcal{S}$, to maximize the system throughput. This approach is concerned with maximizing the number of transmitted bits and serving as many users as possible.
     \item An alternative solution, which is concerned with the total transmit power per frame, is to minimize the transmit power by assigning the time slots to the \emph{frame owners} such that the total transmit power is minimized.
\begin{figure}
  \centering
  \includegraphics[width=0.6\columnwidth]{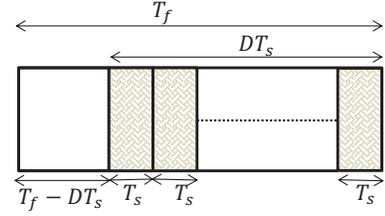}\\
  \caption{Frame structure for the slotted-time data transmission.}
  \label{slotted}
  \end{figure}
Specifically, if the number of decoded requests is less than the sparsity level $\mathcal{S}$, we assume that the destination assigns multiple time slots to some users based on their queue states. Hence, users use lower transmit powers for data transmissions. For this power-saving approach, the nodes can be assigned as follows. Each participating node gets a slot; afterwards, the queues with the highest occupancy among the \emph{frame owners} are assigned to the remaining time slots and so on until the end of all time slots. In this approach, if $\mathcal{T}<D$, the destination announces the value of $\mathcal{T}$, the users' IDs, and the number of time slot assigned to each user. Hence, the communication of the control signals between the destination and the users requires $1/W+2\mathcal{T} /W$ time units. Consequently, the duration of a data time slot is $ T_{\rm s}~=~\max\{\frac{T-{\mathcal{M}/W - 2\mathcal{T}/W-1/W}}{D},0\}$. If $\mathcal{T}=D$, which means that each user will be assigned to only one time slot, the destination does not need to send the number of time slots assigned to each user; hence, the communication of the control signals between the destination and the users requires only $1/W+\mathcal{T} /W$ time units. Consequently, the duration of a data time slot in this case is $ T_{\rm s}~=~\max\{\frac{T-{\mathcal{M}/W - \mathcal{T}/W-1/W}}{D},0\}$.
         \end{enumerate}

%
%

In the second above-mentioned approach, the transmit power of transmitter $k$ when it sends $B_k$ bits over its assigned data time slots, which can be $n_k\in \{1,2,\dots,D\}$ time slots, is given~by
\begin{equation}
\label{gogo111}
P_k=\kappa\left(2^\frac{B_k}{n_k WT_{\rm s}}-1\right)
\end{equation}
The total transmit power per frame is thus given by
\begin{equation}
\label{gogo111}
\sum_{k=1}^\mathcal{N} P_k=\kappa\sum_{k=1}^\mathcal{N} \left(2^\frac{B_k}{n_k WT_{\rm s}}-1\right)
\end{equation}
It is clear that the transmit power is exponentially decreasing with $WT_{\rm s} \frac{n_k}{B_k}$. For a given set of \emph{frame owners}, $W$ and $T_{\rm s}$ are fixed; hence, $\frac{n_k}{B_k}$ is the only term that controls the value of the transmit powers. To maintain the transmit power low, the destination increases the number of time slots assigned to the users with high queues occupancy over the users with low queues occupancy. Thus, we conclude that users with higher queue states must be allocated more time slots than other users. If the number of requests is greater than $\mathcal{S}$, the destination cannot assign more than one time slot to any user. However, if the number of requests is less than $\mathcal{S}$, we can assign more time slots to one or more users based on their buffer states.

To maximize the performances of the proposed schemes, we employ the following algorithm. For each request access probability, $\alpha$, we obtain the sparsity level by solving Eqn. (\ref{sparsity}). Then, for each sparsity level, we choose the minimum cardinality of the observations, i.e., $\mathcal{M}$, such that the probability of users request decoding is close to $1$. Using the resultant sparsity-measurements pairs, we can optimize the system performance efficiently.

In what follows, we investigate the case of full-duplex nodes \cite{choi2010achieving,jain2011practical,bharadia2013full} which enables a time-efficient communication between the users and their destination.
In this case, each node has two radio-frequency (RF) chains and is equipped with one antenna for each RF chain.\footnote{The implementation of full-duplex communications using a single antenna for both data transmission and reception can be performed as in \cite{bharadia2013full}.} Hence, nodes can transmit and receive at the same time. To exploit the ability of nodes to transmit and receive simultaneously, we assume that each node attempts to decode the compressed signal that bears the information about queue states. Thus, each node in the system knows the queue states of the other nodes. Accordingly. we can implement the proposed schemes in a distributed manner. The implementations of the proposed schemes with full-duplex nodes are as follows:
\begin{itemize}
\item For the unslotted-time data transmission scenario, we let transmitter $1$ be the one which uses the spectrum during the time duration from instant $0$ to instant $T_1$, transmitter $2$ is the one which uses the spectrum from instant $T_1$ to instant $T_2$, and so on until the end of the frame. Each transmitter starts and ends its data transmission according to the predefined transmissions sequence.
    \item For the slotted-time data transmission scenario, given that the queue states are known at all nodes, the nodes access the channel based on a predefined order for channel accessing until each transmitter uses only one slot. Afterwards, if the number of \emph{frame owners} is lower than the number of time slots per frame, the assignment of transmitters to the remaining time slots is based on the queues occupancy until the end of the frame. Using these approaches for the unslotted- and slotted-time data transmission scenarios, there is no need for the feedback signals from the destination to the users which consume a large portion of the frame time, especially when the number of users is large.
\end{itemize}

We emphasize here that the residual self-interference caused by the transmission and reception of data at the same time and over the same frequency band does not have any effect on our analysis. That is, let us assume that the self-interference is modeled as a fading coefficient with value $g_k$ where $g_k$ is a circularly-symmetric Gaussian random variable that remains constant during a frame but changes from one frame to another \cite{choi2010achieving,jain2011practical,bharadia2013full}. If the user sends zero, which is the case when its queue is empty or it decides not to access, the signal received at the user due to self-interference is zero. Hence, the self-interference has no impact on the received signal at the user's receiver. If user $k$ sends a request, the received signal at user $k$ is given by
\begin{equation}
\label{for}
\mathbf{Y}^k= \mathcal{A} \mathbf{X}^k+\mathbf{Z}^k
\end{equation}
where $\mathbf{Y}^k$ is the measurement vector received at the $k$th user's receiver, $\mathbf{X}^k=[x_1,x_2,\dots,g_k x_k,\dots,x_\mathcal{N}]^\mathbb{T}$ is the user buffer state vector at the $k$th user, $\mathcal{A}$ is the measurement matrix adopted by the system, and $\mathbf{Z}^k$ is the additive noise vector at the $k$th user's receiver. The OMP output is the same as in the case of no self-interference but the $k$th index is $g_k x_k$ instead of $x_k$. Given that the user knows its own state, i.e., knows that its transmitted signal is $x_k$, it does not need to know the value of $g_k$. Note that all users and the destination solve the same problem in (\ref{for}) with different values of measurements, noise, and residual self-interference coefficient.
 \begin{figure}
  \centering
  \includegraphics[width=1\columnwidth]{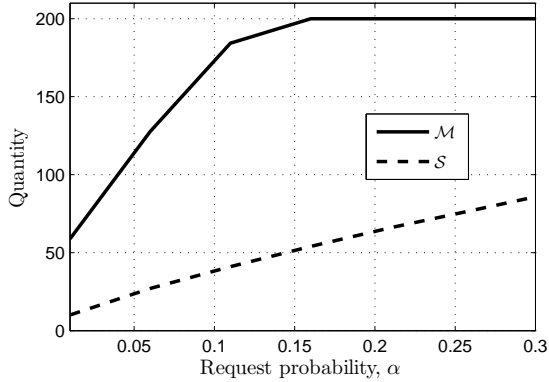}\\
  \caption{The request probability versus $\mathcal{S}$ and $\mathcal{M}$.}
  \label{fig1}
  \end{figure}

\section{Simulations}
\label{sim}
In this section, we evaluate the performances of our proposed schemes. For convenience, we denote the CS-aided schemes for the unslotted-time data transmission and slotted-time data transmission by $\mathcal{P}_{\rm CS}$ and $\hat{\mathcal{P}}_{\rm CS}$, respectively. The case where each user sends its queue state individually for $W/\mathcal{N}$ seconds under the unslotted and slotted scenarios are denoted by $\mathcal{P}$ and $\hat{\mathcal{P}}$, respectively. We model the arrivals at the queues as Bernoulli distributed random variables with mean $\lambda_k=\lambda$ for all queues. For the measurement matrix, we assume that Bernoulli sequences are used by the transmitters.

In Fig. \ref{fig1}, we plot the request probability, $\alpha$, for $\mathcal{N}=200$ users versus both the sparsity level and minimum measurements needed for a given sparsity level when the probability of decoding one of the elements in the request vector is less than $10^{-3}$. The sparsity level, $\mathcal{S}$, is obtained from Eqn. (\ref{sparsity}) with $\epsilon=10^{-4}$. From the figures, and since the sparsity level must be much lower than the measurements length as well as the number of users, the maximum request probability must be less than $0.15$, i.e., $\alpha<0.15$, otherwise, we need $\mathcal{M}=\mathcal{N}$ observations (measurements) to solve the compressed signal. For the parameters used, we find that the ratio between $\mathcal{M}$ and $\mathcal{S}$ is around $5$. Thus, using $c=5$, where $c=\mathcal{M}/\mathcal{S}$, is a very reasonable approximation of the optimal $\mathcal{M}$ for a given sparsity level. This matches the observations in the literature that the number of measurements is empirically given by $c \mathcal{S}$, where $c=\{3,4\}$ \cite{lin2012compressive}.

\begin{figure}
  \centering
  \includegraphics[width=1\columnwidth]{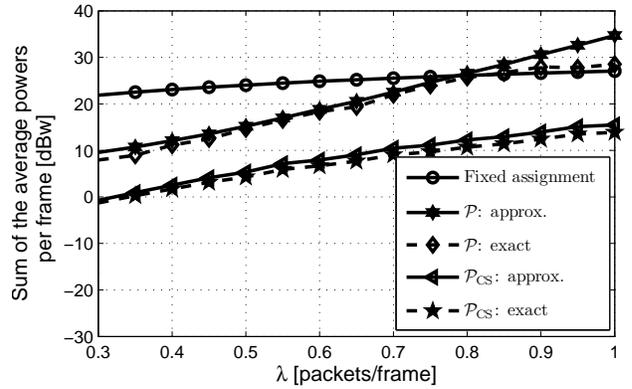}\\
  \caption{Sum of the average powers per frame for the considered schemes.}
  \label{fig5}
  \end{figure}

\begin{figure}
  \centering
  \includegraphics[width=1\columnwidth]{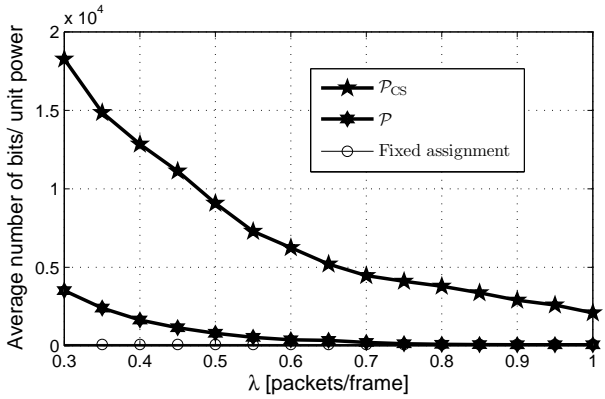}\\
  \caption{Average number of bits per unit power.}
  \label{fig2}
  \end{figure}

\begin{figure}
  \centering
  \includegraphics[width=1\columnwidth]{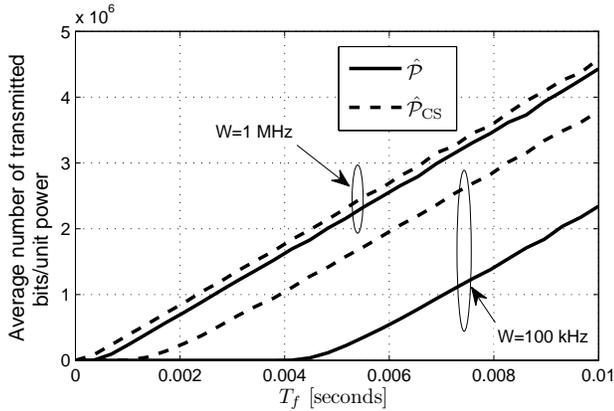}\\
  \caption{Average number of transmitted bits per unit power.}
  \label{fig3}
  \end{figure}

\begin{figure}
  \centering
  \includegraphics[width=1\columnwidth]{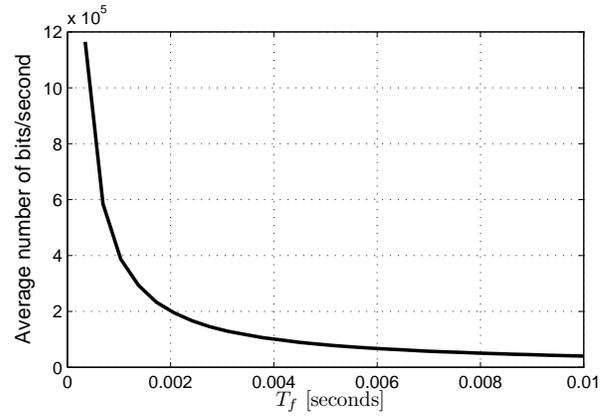}\\
  \caption{Average number of transmitted bits per second.}
  \label{fig4}
  \end{figure}

 Unless otherwise explicitly stated, we assume that Figs. \ref{fig5}, \ref{fig2}, \ref{fig3}, and \ref{fig4} are generated using a probability of decoding one of the elements in the buffer state vector to be less than $10^{-3}$, $\epsilon=10^{-4}$, $\mathcal{N}=400$ users, $\kappa=10^{-9}$ Watts/Hz, maximum buffer size $\mathcal{L}=10$ packets, number of bits per packet $b=100$ bits, $W=5$ MHz, and $T_f=1$ ms.

  The case of unslotted-time data transmission with half-duplex nodes are considered in Figs. \ref{fig5} and \ref{fig2}. Fig. \ref{fig5} demonstrates the impact of the queues arrival rates on the sum of the average powers used by the transmitters. The total power increases with $\lambda$ since the queues occupancy increase and hence the power needed to deliver the bits to the destination. The fixed-assignment scheme has the highest transmit power since each transmitter has only $1/\mathcal{N}$ of the frame duration to transmit its bits. Thus, its bits per unit power
is small. The figure reveals that the analytical solution of our approximate formulation is close to the optimal solution of the total power minimization problem.
%

In Fig. \ref{fig2}, we plot the number of transmitted bits per unit power for the proposed schemes. Our proposed schemes have much higher bits per unit power than the fixed-assignment scheme. The CS-aided scheme, where we send requests and exploit the system sparsity, outperforms the case when each user sends its buffer state individually. The fixed-assignment scheme has the lowest average number of bits/Watts which is $78$ bits/Watts. For high arrival rate at the queues, the sum of the average powers per frame for $\mathcal{P}$ is almost equal to that of the fixed-assignment. This is because at high $\lambda$ almost all queues have the same occupancy and therefore the optimal solution is to assign $1/\mathcal{N}$ of the remaining time in the frame to each buffer. Accordingly, the performance of $\mathcal{P}$ is equal to the fixed-assignment scheme.

Figs. \ref{fig3} and \ref{fig4} study the case of slotted transmissions for full-duplex nodes.
Fig. \ref{fig3} shows the average number of transmitted bits per unit power for the proposed schemes versus the frame duration, $T_f$. The figure considers the case when the system is naturally sparse, i.e., when the arrival rates to the queues are small and the number of users is large. The mean arrival rate of $Q_k$ is $\lambda=0.01$ packets/frame. Since $\lambda=0.01$ packets/frame and $\mathcal{N}=400$, the system is naturally sparse with sparsity level $\mathcal{S}=14$ and the needed measurements is $\mathcal{M}=82$ samples.
 The case of CS-aided scheme outperforms the case when each user sends its queue state separately. The number of bits per seconds under the parameters used in Fig. \ref{fig3} is shown in Fig. \ref{fig4}. We note that as the frame duration increases, the average number of bits per second decreases. This is because, for given queue states detected by the destination, all bits will be delivered to the destination during the frame time regardless of the frame length. The only difference is the transmit power needed to deliver the bits. As the frame length increases, the needed power decreases. For Figs. \ref{fig3} and \ref{fig4}, we assume that the number of data time slots per frame is $D=20$.


\section{Conclusions}
\label{conc}
In this paper, we have proposed new multiple-access schemes for large wireless networks when the nodes are equipped with data buffers. We showed that the system is naturally sparse when the arrival rates to the queues are small and the number of users is large. In this case, we can directly apply the CS sparse signal recovery techniques to obtain the queue states of the transmitters at the destination with few measurements (i.e. signaling time to inform the destination regarding the state of each buffer). We designed the system for the non-sparse case and enforced the user requests to be sparse using the appropriate request probability. For the total transmit power minimization problem, we showed that the problem is convex. Moreover, we proposed an approximate formulation which has a closed-form solution and showed that it achieves a minimum total transmit power per frame close to that of the original optimization problem. Our numerical results demonstrated that the proposed schemes achieve higher average bits per unit power than the fixed-assignment scheme, where each node is assigned to $1/\mathcal{N}$ of the frame duration. In addition, the average number of bits/unit power increases with increasing the frame time and/or transmit bandwidth. However, the number of delivered bits per unit time to the destination decreases with increasing frame duration, $T_f$.

\bibliographystyle{IEEEtran}
\bibliography{IEEEabrv,term_bib}
\balance
\end{document}